\documentclass[%
 reprint,
 superscriptaddress,
 amsmath,amssymb,
 aps,
]{revtex4-2}

\usepackage[english]{babel}
\usepackage{graphicx}
\usepackage{dcolumn}
\usepackage{bm}
\usepackage[caption=false]{subfig}
\usepackage{color}
\usepackage{ulem}
\usepackage{siunitx}

\newcommand{\secref}[1]{Sec.~\ref{#1}}

\newcommand{\Mc}{\mathcal{M}}

\begin{document}

\preprint{}

\title{Mass--spin Re-Parameterization for Rapid Parameter Estimation\\ of Inspiral Gravitational-Wave Signals}

\author{Eunsub Lee}
 \affiliation{Institute for Cosmic Ray Research, The University of Tokyo,
5-1-5 Kashiwanoha, Kashiwa, Chiba 277-8582, Japan}
\author{Soichiro Morisaki}
 \affiliation{Department of Physics, University of Wisconsin-Milwaukee, Milwaukee, WI 53201, USA}
\author{Hideyuki Tagoshi}
 \affiliation{Institute for Cosmic Ray Research, The University of Tokyo,
5-1-5 Kashiwanoha, Kashiwa, Chiba 277-8582, Japan}

\date{\today}

\begin{abstract}
Estimating the source parameters of gravitational waves from compact binary coalescence(CBC) is a key analysis task in gravitational-wave astronomy.
To deal with the increasing detection rate of CBC signals, optimizing the parameter estimation analysis is crucial.
The analysis typically employs a stochastic sampling technique such as Markov Chain Monte Carlo(MCMC), where the source parameter space is explored and regions of high Bayesian posterior probability density are found.
One of the bottlenecks slowing down the analysis is the non-trivial correlation between masses and spins of colliding objects, which makes the exploration of mass--spin space extremely inefficient.
We introduce a new set of mass--spin sampling parameters which makes the posterior distribution to be simple in the new parameter space, regardless of the true values of the parameters.
The new parameter combinations are obtained as the principal components of the Fisher matrix for the restricted 1.5 post-Newtonian waveform.
Our re-parameterization improves the efficiency of MCMC by a factor of $\sim10$ for binary neutron star with narrow-spin prior ($|\vec{\chi}|<0.05$) and $\sim100$ with broad-spin prior ($|\vec{\chi}|<0.99$), under the assumption that the binary has spins aligned with its orbital angular momentum.
\end{abstract}

\maketitle


\section{Introduction} \label{sec:introduction}
In September 2015, the first direct detection of gravitational waves(GWs) took place, which was radiated from the coalescence of a binary black hole(BBH) \cite{PhysRevLett.116.061102}.
Since the first detection, tens of gravitational-wave signals from compact binary coalescence(CBC) have been reported by the LIGO--Virgo--KAGRA collaboration \cite{Abbott_2019, Abbott:2020niy, LIGOScientific:2021usb, LIGOScientific:2021djp}, including the coalescence of a binary neutron star(BNS) with electromagnetic(EM) follow-up observations \cite{Abbott_2017}.  

Estimating source parameters from a CBC signal is an important task in gravitational-wave astronomy.
The estimated source location is crucial for the EM follow-up observations, and the masses and spins of colliding objects are important for studying the formation history of compact binaries \cite{LIGOScientific:2020kqk, LIGOScientific:2021psn}.

This parameter estimation analysis typically employs Bayesian inference using stochastic sampling techniques, such as Markov Chain Monte Carlo(MCMC) \cite{doi:10.1063/1.1699114,10.1093/biomet/57.1.97} and Nested sampling \cite{Skilling2006}.
While the stochastic sampling is known to be efficient for estimating high dimensional parameters, it is still computationally costly, taking more than weeks for a BNS event without any approximate methods.
Speeding it up is necessary to deal with the increasing detection rate of CBC signals.
It is also crucial for the follow-up observations of EM counterparts rapidly fading after the merger.

Efficient exploration of the parameter space is essential for optimizing the stochastic sampling. One-dimensional jumps can efficiently explore the parameter space if parameters are not strongly correlated. However, they are extremely inefficient if parameters are strongly correlated, which is the case for a CBC signal. 
One solution for this is to use parameterizations minimizing the correlations between parameters. Based on this idea, the \texttt{LALInference} software \cite{Veitch:2014wba} uses chirp mass $\Mc \equiv (m_1 m_2)^{3/5} (m_1 + m_2)^{-1/5}$  and mass ratio $q \equiv m_2 / m_1$ to reduce the correlation between component masses, $m_1$ and $m_2$, where $m_1 \geq m_2$. However, masses 
are correlated with spins \cite{Cutler:1994ys}, and this choice of parameters does not fully minimize their correlations.

In this paper, we introduce an alternative set of mass--spin parameters which has significantly reduced correlations between parameters.
The new parameters, referred to as $\mu_1$ and $\mu_2$, have been studied in literature for efficient template placement \cite{Tanaka_2000,Sathyaprakash_2003,Brown_2012} and to find the best measurable combinations of physical parameters \cite{Ohme_2013,Pai_2012,Morisaki_2020}. In contrast to them, we use these parameters in the sampling process in MCMC. We show that using the new parameterizations significantly speeds up the parameter estimation, without the loss of accuracy of the estimation.

This paper is organized as follows.
In \secref{sec:background}, we introduce the basics of parameter estimation and the correlation between mass and spin parameters.
In \secref{sec:methodology}, we introduce a new set of mass--spin parameters, which has significantly reduced correlations between parameters, and explain a sampling method using them.
In \secref{sec:injection_test}, we describe injection tests for studying the performance of our new sampling strategy and introduce their results.
Finally we present our conclusion in \secref{sec:conclusions}.

\section{Background} \label{sec:background}

In this section, we explain the basics of Bayesian parameter estimation using MCMC. Then, we introduce the correlation between mass and spin parameters, which makes the parameter estimation analysis of a CBC signal significantly inefficient.

\subsection{Parameter estimation using MCMC} 
In the Bayesian inference, the inference result is the posterior distribution $p(\theta|d)$ which is the probability distribution of the parameters $\theta$ when the observation data $d$ is given.
The posterior distribution is calculated by the Bayes' theorem,
\begin{equation}
\label{eqn:posterior}
    p(\theta|d) \propto p(d|\theta)p(\theta),
\end{equation}
where $p(d|\theta)$ is the likelihood at $\theta$ and $p(\theta)$ is the prior distribution.
We assume that the data of a detector is modeled as the sum of Gaussian, stationary random noise $n$ and a gravitational-wave signal $h$,
\begin{equation}
    d(t) = n(t) + h(t;\theta).
\end{equation}
Then the likelihood for a single detector becomes
\begin{equation}
\label{eqn:likelihood}
        p(d|\theta)
        \propto \exp\left[-\frac{1}{2}
        (d-h(\theta),d-h(\theta))\right].
\end{equation}
Here $(x, y)$ indicates the noise-weighted inner product,

\begin{equation}
(x, y) \equiv 4 \Re \left[ \int_{f_{\mathrm{min}}}^{f_{\mathrm{max}}} df \frac{\tilde{x}^\ast(f) \tilde{y}(f)}{S_n(f)} \right],
\end{equation} where $f_{\mathrm{min}}$ and $f_{\mathrm{max}}$ are the low- and high-frequency cutoffs of the analysis respectively, and $S_n(f)$ is the one-sided power spectral density(PSD) of the detector. For multiple detectors, the likelihood of the combined data is the product of that of each detector, assuming the noise at each detector is statistically independent.

Even though the posterior distribution has a simple form of Eq.(\ref{eqn:posterior}), it is not easy to get the properties of this distribution.
A practical method is to generate samples that follow the distribution.
MCMC is an efficient method for the sampling from the posterior distribution.
A new sample is drawn stochastically based on the current sample.
The most basic MCMC is the Metropolis-Hastings(MH) algorithm\cite{10.1093/biomet/57.1.97}.
In MH algorithm, drawing a sample is divided into two steps: proposal and acceptance-rejection. Selecting appropriate proposal distribution is critical to the performance of MH algorithm. However, if the posterior distribution has complicated structure owing to the correlation of parameters, it is a hard task to find efficient proposal distribution in advance.

\begin{figure}[t]
\centering
\includegraphics[width=\linewidth]{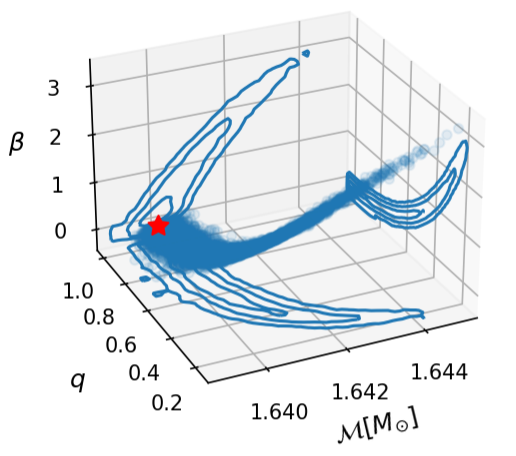}
\caption{Posterior samples from a simulated signal, which is presented as \#S2 in \secref{sec:injection_test}. 2-dimensional plots are marginal distributions and contours represent $1\sigma$, $2\sigma$ and $3\sigma$ regions. The red star marks the true values of the parameters.}
\label{fig:3d}
\end{figure}

\subsection{Correlation between masses and spins}

The phase evolution of gravitational waves is predominantly characterized by the masses and spins of colliding objects.
Especially, the leading term in the post-Newtonian expansion solely depends on chirp mass $\Mc$.
Thus, the chirp mass is precisely determined from gravitational waves, and the contour of mass distribution approximately follows the fixed line of $\Mc$. 
This leads to the strong correlation between $m_1$ and $m_2$, which makes the stochastic sampling in the $m_1$--$m_2$ coordinate system inefficient.
As explained in \secref{sec:introduction}, the LALInference software uses $\Mc$ and mass ratio $q$ as independent sampling parameters to solve this issue.

The dominant spin contribution appears at the $1.5$ post-Newtonian order through the following combination,
\begin{equation}
\beta = \frac{1}{12}\sum_{i=1}^{2} \left[113\left(\frac{m_i}{m_1+m_2}\right)^2+75\eta\right]{\chi}_i, \label{eq:beta}
\end{equation}
where $\eta$ is symmetric mass ratio,
\begin{equation}
\eta = \frac{m_1 m_2}{(m_1 + m_2)^2},
\end{equation}
and $\chi_i$ is the component of the spin angular momentum $\vec{S}_i$ along the orbital angular momentum $\vec{L}_i$ normalized by $m^2_i$,
\begin{equation}
\chi_i \equiv \frac{\mathbf{L}}{\left| \mathbf{L} \right|} \cdot \frac{c \mathbf{S}_i}{G m^2_i}.
\end{equation}
Since it also affects the frequency evolution, it is correlated with $\Mc$ and $q$.
Figure~\ref{fig:3d} shows the posterior samples for a binary neutron star signal in the $\Mc$--$q$--$\beta$ space.
It shows that the mass and spin parameters are correlated non-trivially. Even $\Mc$ and $q$ are strongly correlated with each other again in this case. The samples are along a characteristic curve, thus the mass--spin space needs to be explored along the curve, which makes the sampling difficult and inefficient.

\section{Methodology} \label{sec:methodology}

In this section, we construct an alternative set of mass--spin sampling parameters, which has significantly reduced correlation between parameters.
Following \cite{Ohme_2013, Morisaki_2020}, we construct them as the principal components of the Fisher matrix for the restricted post-Newtonian waveform.
We also discuss the practical choice of sampling parameters, sampling method and its extension to the case of multiple detectors.

\subsection{Restricted post-Newtonian waveform}
To construct the efficient sampling parameters, we make use of a restricted post-Newtonian waveform model. Here we take into account terms up to the 1.5 post-Newtonian order to incorporate the dominant spin contribution. 
The waveform is given by
\begin{equation}
\label{eqn:waveform}
    \tilde{h}(f) = \mathcal{A}\left(\frac{f}{f_\mathrm{ref}}\right)^{-\frac{7}{6}}e^{-i\Psi(f)},
\end{equation}
with the phase function
\begin{eqnarray}
    \Psi(f) =\;&&  
    \psi_1\left(\frac{f}{f_\mathrm{ref}}\right)^{-\frac{5}{3}}
    +\psi_2\left(\frac{f}{f_\mathrm{ref}}\right)^{-1}
    +\psi_3\left(\frac{f}{f_\mathrm{ref}}\right)^{-\frac{2}{3}}\nonumber\\
    &&+\psi_4
    +\psi_5\left(\frac{f}{f_\mathrm{ref}}\right).
\end{eqnarray}
The phase expansion coefficients are related to the physical parameters as
\begin{subequations}
\begin{align}
\label{eqn:psis}
    \psi_1 &= \frac{3}{128}(\pi G\mathcal{M}f_\mathrm{ref}/c^3)^{-\frac{5}{3}},\\
    \psi_2 &= \frac{55}{384}\left(\eta+\frac{743}{924}\right)\eta^{-\frac{2}{5}}(\pi G\mathcal{M}f_\mathrm{ref}/c^3)^{-1},\\
    \psi_3 &= \frac{3}{32}(\beta - 4\pi)\eta^{-\frac{3}{5}}(\pi G\mathcal{M}f_\mathrm{ref}/c^3)^{-\frac{2}{3}},\\
    \psi_4 &= -2\phi_c-\frac{\pi}{4},\\
    \psi_5 &= 2\pi f_\mathrm{ref} t_c,
\end{align}
\end{subequations} where $\phi_c$ is merger phase, $t_c$ is merger time and $\beta$ is the combination of spins defined as \eqref{eq:beta}. 
$f_\mathrm{ref}$ is a reference frequency introduced to make phase expansion coefficients dimensionless, and we use $f_\mathrm{ref}=200\,\si{\hertz}$ following \cite{Ohme_2013, Morisaki_2020}.
The amplitude $\mathcal{A}$ in Eq.(\ref{eqn:waveform}) is a function of the chirp mass and the extrinsic parameters, and the signal-to-noise ratio(SNR) $\varrho \equiv (h,h)^{-1/2}$ is proportional to $\mathcal{A}$.

\subsection{Principal component analysis}

We use the restricted post-Newtonian waveform to study the approximate structure of the posterior distribution in the mass--spin space.
By substituting the restricted 1.5P post-Newtonian waveform into the likelihood for a single detector \eqref{eqn:likelihood}, we obtain
\begin{equation}
\label{eqn:likelihoodpsi}
\begin{split}
        p(d|\psi) \propto
        \exp\Bigl[&-\frac{1}{2}(h(\hat{\psi})-h(\psi),h(\hat{\psi})-h(\psi))\\
         &-(n,h(\hat{\psi})-h(\psi))
        \Bigr],
\end{split}
\end{equation}
where we use the phase coefficients $\psi=\{\psi_i\}_{i=1}^5$ instead of physical parameters to parameterize the waveform, and $\hat{\psi}_i$ is their true values.
In the limit of a high signal-to-noise ratio, posterior distribution has a sharp peak around $\hat{\psi}$. We expand $h(\psi)$ as a Taylor series around $h(\hat{\psi})$ and approximate it to the leading order. Then Eq.(\ref{eqn:likelihoodpsi}) becomes Gaussian,
\begin{equation}
\label{eqn:fisher}
p(d|\psi) \propto
        \exp\left[ -\frac{1}{2}\sum_{i,j}\Gamma_{ij}\Delta\psi_i\Delta\psi_j\right],
\end{equation}
where 
\begin{equation}
    \Gamma_{ij} \equiv \left(
        \frac{\partial h}{\partial \psi_i},\frac{\partial h}{\partial \psi_j}
        \right)
\end{equation}
is the Fisher information matrix(FIM) for $\psi$, and
\begin{equation}
\Delta \psi_i \equiv \psi_i - \hat{\psi}_i - \sum_k \left(\Gamma^{-1}\right)^{ik} \left(n, \frac{\partial h}{\partial \psi_k} \right).
\end{equation}

We construct combinations which depend on  $(\mathcal{M},q,\chi_1,\chi_2)$ but not $t_c$ and $\phi_c$, in consideration of the easy availability. We thus consider the posterior marginalized over $\phi_c$ and $t_c$, or equivalently $\psi_4$ and $\psi_5$. The marginal posterior is then 
\begin{equation}
\label{eqn:posteriorpsi}
\begin{split}
        &p(\{\psi_1,\psi_2,\psi_3\}|d)\\
        \propto\; &p(\{\psi_1,\psi_2,\psi_3\})\exp \left[-\frac{1}{2}\sum_{i,j}\tilde{\Gamma}_{ij}\Delta\psi_i\Delta\psi_j\right],
\end{split}
\end{equation}
where $p(\{\psi_1,\psi_2,\psi_3\})$ is the prior distribution of $(\psi_1, \psi_2, \psi_3)$, $\tilde{\Gamma}$ is the 3$\times$3 FIM,
\begin{equation}
    \tilde{\Gamma}_{ij} = \Gamma_{ij} - \sum_{k,l}\Gamma_{ik}\gamma^{kl}\Gamma_{lj},
\end{equation}
$i,j =1,2,3$, $k,l = 4,5$ and 
\begin{equation}
    \gamma =\begin{bmatrix}
    \Gamma_{44} & \Gamma_{45}\\
    \Gamma_{54} & \Gamma_{55}
    \end{bmatrix}
    ^{-1}.
\end{equation}

We can diagonalize $\tilde{\Gamma}$ using an orthogonal matrix $U$ as
\begin{equation}
    \tilde{\Gamma}_{ij}
    = \sum_{m,n}U^\mathrm{T}_{im}\lambda_m\delta_{mn}U_{nj},
\end{equation}
where $\{\lambda_m\}_{m=1}^3$ are the eigenvalues in descending order $\lambda_1 > \lambda_2 > \lambda_3$.
Using this Eq.(\ref{eqn:posteriorpsi}) takes a simple form,
\begin{equation}
\label{eqn:posteriormu}
        p(\mu|d) \propto p(\mu)\prod_n e^{-\frac{1}{2}\lambda_n \Delta \mu_n^2},
\end{equation}
with a new set of parameters
\begin{equation}
\label{eqn:mu}
    \phantom{(n = 1,2,3)}
    \,\,\,\,\,\,\,\,\,\,\,\,\,\,\,
    \mu_n \equiv \sum_i {U}_{ni }\psi_i, 
    \,\,\,\,\,\,\,\, (n = 1,2,3).
\end{equation}

Eq.(\ref{eqn:posteriormu}) can be represented as product of each parameter's function if the prior distribution $p(\mu)$ is separable. This implies that if the parameters are not strongly correlated in the prior, the posterior distribution in the $\mu$ space becomes very simple. Thus, we use $\mu_n$ as an alternative sampling parameter.

Before discussing the sampling method using $\mu_n$, we discuss the dependence of $\mu_n$ on masses and spins.
The original FIM is given as
\begin{equation}
\label{eqn:fishermatrix}
\Gamma = 4|\mathcal{A}|^2
    \begin{bmatrix}
    I_{-17/3} & I_{-5} & I_{-14/3} & I_{-4} & I_{-3}\\
    I_{-5} & I_{-13/3} & I_{-4} & I_{-10/3} & I_{-7/3}\\
    I_{-14/3} & I_{-4} & I_{-11/3} & I_{-3} & I_{-2}\\
    I_{-4} & I_{-10/3} & I_{-3} & I_{-7/3} & I_{-4/3}\\
    I_{-3} & I_{-7/3} & I_{-2} & I_{-4/3} & I_{-1/3}
    \end{bmatrix},
\end{equation}
with 
\begin{equation}
    I_a \equiv \int_{f_\mathrm{min}}^{f_\mathrm{max}}
        \left(\frac{f}{f_\mathrm{ref}}\right)^a\frac{df}{S_n(f)}.
\end{equation}

$f_\mathrm{max}$ is usually set to the innermost stable circular orbit frequency,
\begin{equation}
    f_{\mathrm{isco}} = \frac{1}{6^{3/2}\pi G(m_1+m_2)/c^3},
\end{equation} 
for inspiral-only waveform models. 
Thus, $\Gamma$, and hence $\mu_n$, generally depends on masses. However, in this work, we fix $f_\mathrm{max}$ to $2048\,\si{\hertz}$ instead.
Then $U$ does not depend on the physical parameters, and depends only on  $f_{\mathrm{min}}$ and the power spectral density of the detector. 
For typical BNS events, $f_\mathrm{isco} \simeq 4400/(M/M_\odot)\si{\hertz}$ is higher than $1000\,\si{\hertz}$ for BNS events. Since $S_n(f)$ gets larger in $f > 1000\,\si{\hertz}$ for the current ground-based detectors, and $(f / f_\mathrm{ref})^\alpha$ simply decreases for $\alpha<0$, the contributions to $I_\alpha$ for $2048\,\si{\hertz} < f < f_\mathrm{isco}$ or $f_\mathrm{isco} < f < 2048\,\si{\hertz}$ are suppressed.
Thus, our choice of $f_{\mathrm{max}}$ is a reasonable approximation for BNS events. Actually, even if we change $f_\mathrm{max}$ to $1000\,\si{\hertz}$,
the coefficients of $\mu_1$ and $\mu_2$ are changed up to 5.6\% compared to when $f_\mathrm{max} = 2048$.
In \secref{sec:injection_test}, we show that $\mu_n$ obtained under this choice of $f_{\mathrm{max}}$ is useful even for BBH events.

\subsection{Practical choice of sampling parameters and sampling strategy}

Next, we discuss our choice of sampling parameters including $\mu_n$.
In this paper, we restrict ourself to a system whose spins are aligned with the orbital angular momentum and consider only 4 mass--spin parameters $(m_1, m_2, \chi_1, \chi_2)$.
We will discuss a potential extension of our method to the case of full spin components in \secref{sec:conclusions}.

As a surrogate set of mass--spin parameters, we choose $(\mu_1,\mu_2,q,\chi_2)$. 
The prior distribution in $(\mu_1,\mu_2,q,\chi_2)$ space can be represented as
\begin{equation}
    p(\mu_1,\mu_2,q,\chi_2) = \frac{p(m_1,m_2,\chi_1,\chi_2)}{|J|}.
\end{equation}
$|J|$ is the Jacobian determinant, which can be calculated as
\begin{equation}
\begin{split}
    |J|= &\left|(U_{11}U_{23}-U_{21}U_{13})\frac{\partial\psi_1}{\partial\mathcal{M}}\frac{\partial\psi_3}{\partial\chi_1}\right.\\
    &\left.+(U_{12}U_{23}-U_{22}U_{13})\frac{\partial\psi_2}{\partial\mathcal{M}}\frac{\partial\psi_3}{\partial\chi_1}
    \right|
    \left| \frac{\mathcal{M}}{m_1^2}\right|,
\end{split} \label{eqn:jacobian}
\end{equation}
where 
\begin{align}
    \frac{\partial\psi_1}{\partial\mathcal{M}} &= 
    -\frac{5}{128}\frac{\pi Gf_\mathrm{ref}}{c^3}(\pi G \mathcal{M} f_\mathrm{ref}/c^3)^{-\frac{8}{3}},\\
    \frac{\partial\psi_2}{\partial\mathcal{M}} &=
    -\frac{55}{384}\left(\eta+\frac{743}{924}\right)\eta^{-\frac{2}{5}}\frac{\pi Gf_\mathrm{ref}}{c^3}(\pi G\mathcal{M}f_\mathrm{ref}/c^3)^{-2},\\
    \frac{\partial\psi_3}{\partial\chi_1} &=
    \frac{113+75q}{12(1+q)^2}\frac{3}{32}\eta^{-\frac{3}{5}}(\pi G\mathcal{M}f_\mathrm{ref}/c^3)^{-\frac{2}{3}}.
\end{align}
To include $q$ is necessary to avoid the singularity of the Jacobian at $m_1=m_2$, which is an obstacle for the use of $\eta$ as a sampling parameter \cite{Veitch:2014wba}. A natural candidate for the sampling parameters includeing $q$ might be $(\mu_1,\mu_2,\mu_3,q)$. However, in that case the Jacobian determinant becomes 0, since $(\mu_1,\mu_2,\mu_3)$ depend on $\chi_1$ and $\chi_2$ only via $\psi_3$.

We can generate posterior samples efficiently in $(\mu_1,\mu_2,q,\chi_2)$. To obtain the posterior distribution for physical parameters, we need transformation from $(\mu_1,\mu_2,q,\chi_2)$ to $(m_1,m_2,\chi_1,\chi_2)$.
$m_1$ and $m_2$ are calculated as follows.
First, $x \equiv \mu_1 - (U_{13}/U_{23}) \mu_2$, is calculated.
Then $\mathcal{M}$ is calculated as a solution of the following equation,
\begin{equation}
\left(U_{11} - \frac{U_{13}}{U_{23}} U_{21}\right) \psi_1 + \left(U_{12} - \frac{U_{13}}{U_{23}} U_{22}\right) \psi_2 = x. \label{eq:bisection}
\end{equation}
Since the left hand side of \eqref{eq:bisection} is a decreasing function of $\mathcal{M}$, we can use a simple bisectional search to find the solution.
$m_1$ and $m_2$ are easily calculated from $\mathcal{M}$ and q.
Finally, given $(\mu_2,m_1,m_2,\chi_2)$, $\chi_1$ is easily calculated.

\subsection{Multiple detector case}
We can also construct $\mu$ parameters satisfying Eq.(\ref{eqn:posteriormu}) when we have data from multiple detectors. For instance, suppose we have 3 detectors, LIGO-Livingston(L), LIGO-Hanford(H) and Virgo(V), and the Fisher matrix for the detectors are $\Gamma_\text{L}$, $\Gamma_\text{H}$ and $\Gamma_\text{V}$ respectively, the likelihood becomes
\begin{equation}
    p(d|\psi) \propto \exp\left[-\frac{1}{2}\Gamma_\mathrm{LHV}\Delta\psi_i\Delta\psi_j\right],
\end{equation}
where $\Gamma_\mathrm{LHV} \equiv \Gamma_\text{L}+\Gamma_\text{H}+\Gamma_\text{V}$.
The Fisher matrix for each detector is proportional to the square of the SNR of the signal at the detector, so we write
\begin{equation}
    \Gamma_\mathrm{LHV} = \varrho^2_\mathrm{L}\hat{\Gamma}_\mathrm{L}+\varrho^2_\mathrm{H}\hat{\Gamma}_\mathrm{H}+\varrho^2_\mathrm{V}\hat{\Gamma}_\mathrm{V},
\end{equation}
where 
\begin{equation}
\label{eqn:fishermatrixhat}
\hat{\Gamma}_\text{L} = 
    \begin{bmatrix}
    I_{-17/3} & I_{-5} & I_{-14/3} & I_{-4} & I_{-3}\\
    I_{-5} & I_{-13/3} & I_{-4} & I_{-10/3} & I_{-7/3}\\
    I_{-14/3} & I_{-4} & I_{-11/3} & I_{-3} & I_{-2}\\
    I_{-4} & I_{-10/3} & I_{-3} & I_{-7/3} & I_{-4/3}\\
    I_{-3} & I_{-7/3} & I_{-2} & I_{-4/3} & I_{-1/3}
    \end{bmatrix}_\text{L}
    \left(I_{-7/3}\right)^{-1}_\text{L},
\end{equation} is parameter independent and therefore can be calculated in advance with the PSD of the detector L, and similarly for H and V. 

Using the point estimate of SNRs given by the detection pipeline, we can calculate $\Gamma_\mathrm{LHV}$, $\tilde{\Gamma}_\mathrm{LHV}$ and ${U}_\mathrm{LHV}$ quickly.
Note that if the PSDs for detectors are very similar we can just use ${U}_{ij}$ calculated from any single detector regardless of the SNRs.

\section{Injection test} \label{sec:injection_test}

\begin{table}[b]
\centering
\begin{ruledtabular}
\begin{tabular}{l l r r  r r}
Spin prior range & Case & $\Mc[M_\odot]$& $q$  & $\chi_1$& $\chi_2$ \\
\hline
\multicolumn{2}{l}{\textbf{Single detector cases}}&&&&\\[1.5ex]
Narrow & \#S1 & 1.64 & 1.0 &  0.02 & 0.02\\
$|\chi_1|,|\chi_2|<0.05$ &  & & & &\\[1.5ex]
Semi-broad& \#S2 & 1.64&1.0&0 &0 \\
$|\chi_1|,|\chi_2|<0.4$ & \#S3 & 1.64&0.7&0 & 0 \\
& \#S4 & 1.64& 1.0& 0.3& 0.3  \\
& \#S5 & 1.64 &0.7 &0.3 &0.3   \\[1.5ex]
Broad & \#S6 &1.64 & 0.7 &0.5 &0.5 \\
$|\chi_1|,|\chi_2|<0.99$ & \#S7 & 28.2 & 0.84 & 0.26 &0. 32\\[1.5ex]
\hline
\multicolumn{2}{l}{\textbf{Multiple detector case}}  &  & & &\\[1.5ex]
$|\chi_1|,|\chi_2|<0.4$& \#M &1.64 & 0.9 &0 &0\\
\end{tabular}
\end{ruledtabular}
\caption{Test cases.}
\label{tab:injectionbinary}
\end{table}

\begin{table}[t]
    \centering
    \begin{ruledtabular}
    \begin{tabular}{l c c c c c}
        Case & Minimum & & Parameter & & Maximum \\
        \hline
        \#S1 & 148.615862 & $\le$ & $\mu_1$ & $\le$ &156.940475\\
        & $-76.106636$ & $\le$ & $\mu_2$ & $\le$ &$-41.460751$\\[1.5ex]
        \#S2--S5 & 145.212081 & $\le$ & $\mu_1$ & $\le$ &158.693980\\
        & $-97.311606$ & $\le$ & $\mu_2$ & $\le$ &$-30.738354$\\[1.5ex]
        \#S6 & 139.474278 & $\le$ & $\mu_1$ & $\le$ &162.409836\\
        & $-133.057127$ & $\le$ & $\mu_2$ & $\le$ &$-12.663454$\\[1.5ex]
        \#S7 & -1.520207 & $\le$ & $\mu_1$ & $\le$ &1.834994\\
        & $-20.906248$ & $\le$ & $\mu_2$ & $\le$ &$-0.742144$\\[1.5ex]
        \#M & 145.231364 & $\le$ & $\mu_1$ & $\le$ &158.702923\\
        & $-96.789911$ & $\le$ & $\mu_2$ & $\le$ &$-30.549108$\\
    \end{tabular}
    \end{ruledtabular}
    \caption{Prior range of $\mu_1$ and $\mu_2$ parameters used in each case.}
    \label{tab:masspriorrange}
\end{table}

To confirm the effectiveness of the new sampling parameters, we conduct a series of injection tests, where CBC signals are artificially injected into simulated Gaussian noise and their source parameter values are recovered by parameter estimation analyses.
The analyses are performed under the assumption that the sources have spins aligned with their angular momenta, and only the two spin components $(\chi_1, \chi_2)$ are sampled.
They are performed with the conventional set of sampling parameters $(\Mc, q, \chi_1, \chi_2)$ and our new set of sampling parameters $(\mu_1, \mu_2, q, \chi_2)$, and their efficiency and estimation results are compared.

The signal injection and parameter estimation process are managed through BILBY\cite{2019ApJS..241...27A}.
The injected CBC signals are generated by the \textsc{IMRPhenomD} waveform model \cite{Husa_2016,Khan_2016}, and the same waveform model is used for recovering the source parameter values.
The \textsc{IMRPhenomD} describes not only inspiral phase, but also merger and ringdown phases in contrast with the PN waveform we used in constructing parameters. 
The waveform calculations are done via \texttt{LALSimulation} \cite{lalsuite}. 
The integration range of the likelihood is from $f_{\mathrm{min}}=20\,\si{\hertz}$ to $f_{\mathrm{max}}=2048\,\si{\hertz}$.

For parameter estimation, we use \texttt{PTMCMCSampler}\cite{justin_ellis_2017_1037579}. Parallel tempering(PT)\cite{PhysRevLett.57.2607,Geyer1991} is the main feature of this sampler and it makes the sampling efficient especially when the probability distribution is multi-modal. Though PT can reduce the auto-correlation of samples, it is not used in our tests since we are more interested in the convergence than searching modes. For the same reason, we fix the starting point of the sampling to the injected parameter values.

\texttt{PTMCMCSampler} provides several built-in jump proposals, and custom proposals can be added to the sampling process. In here, two built-in proposals, Single Component Adaptive Metropolis(SCAM) and Adaptive Metropolis(AM) are used with the same weights. We use default options, with a minor modification: If there is no accepted proposal until the adaptation stage, the scale of jump proposals is changed by $1/2$.

\subsection{Test cases}

\begin{figure*}[ht]
\subfloat{%
  \includegraphics[width=.49\linewidth]{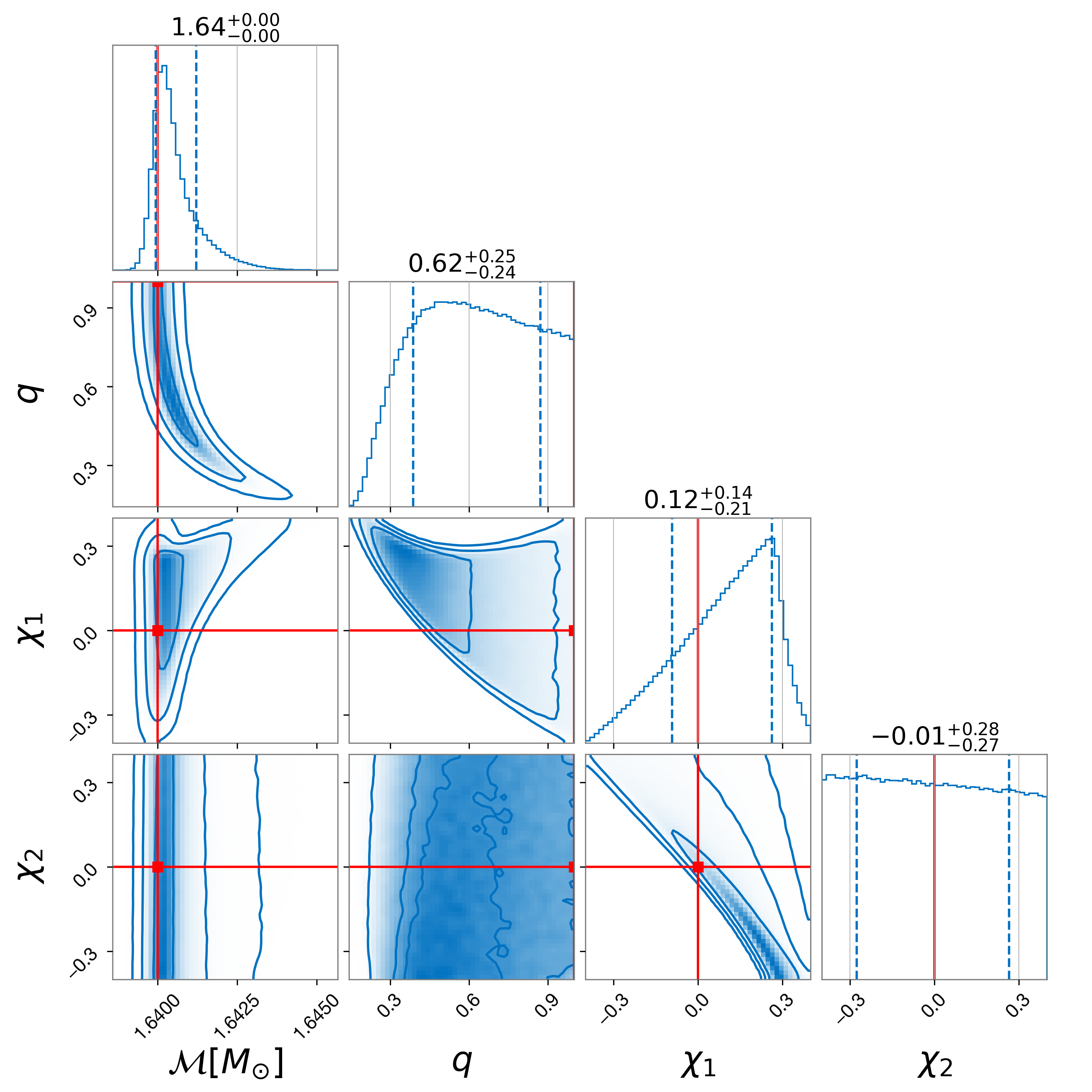}%
}\hfill
\subfloat{%
  \includegraphics[width=.49\linewidth]{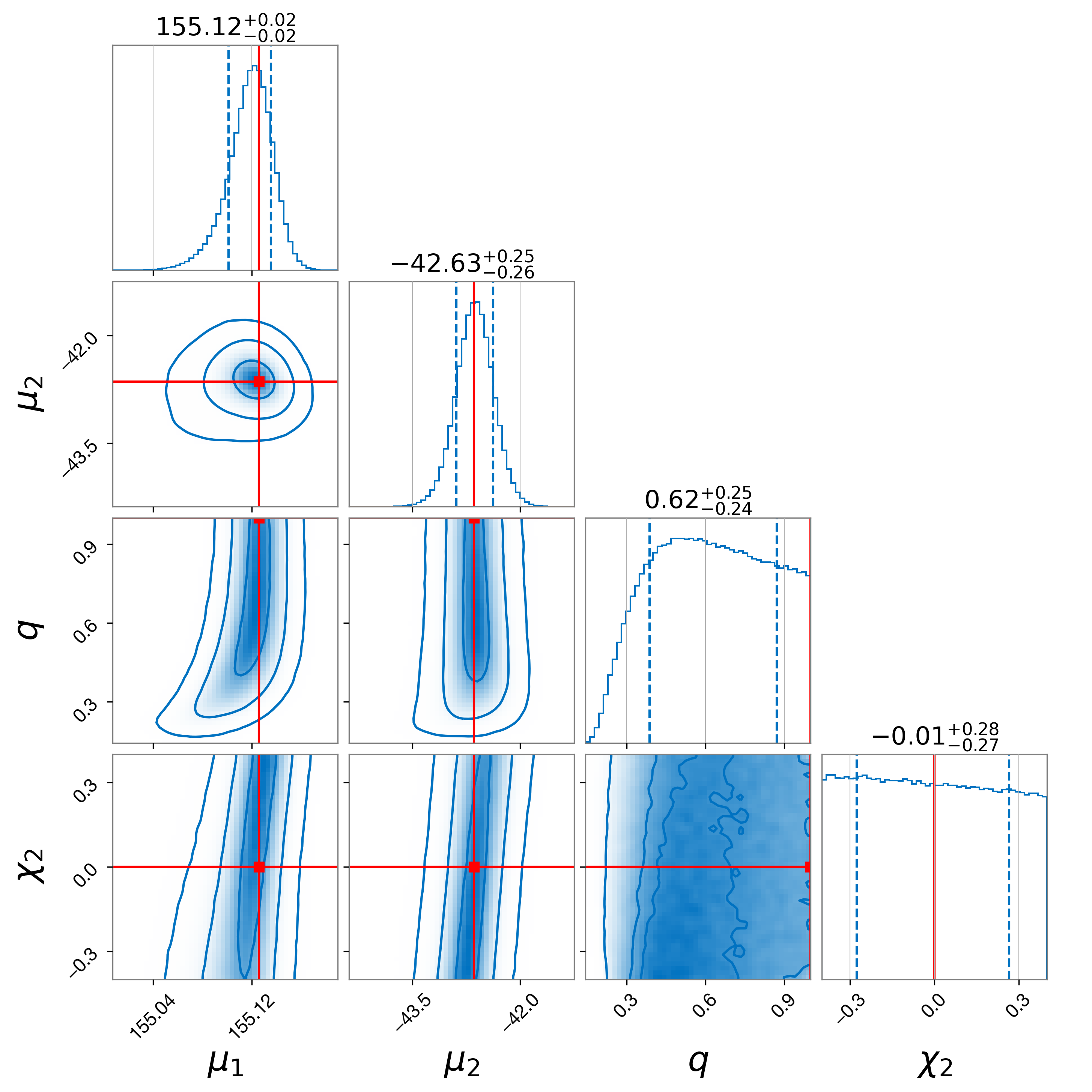}%
}
\caption{Mass--spin part of the estimated posterior distribution in the injection test case \#S2. The same posterior samples are plotted in the two different sets of mass and spin parameters.
Dashed lines on each 1-dimensional marginal distribution represent $1\sigma$ region. 3 contours on each 2-dimensional marginal distribution represent $1\sigma$, $2\sigma$ and $3\sigma$ region respectively. The red lines represent the injected value.}
\label{fig:massspin_test}
\end{figure*}

\begin{figure*}[t]
\centering
\includegraphics[width=0.7\linewidth]{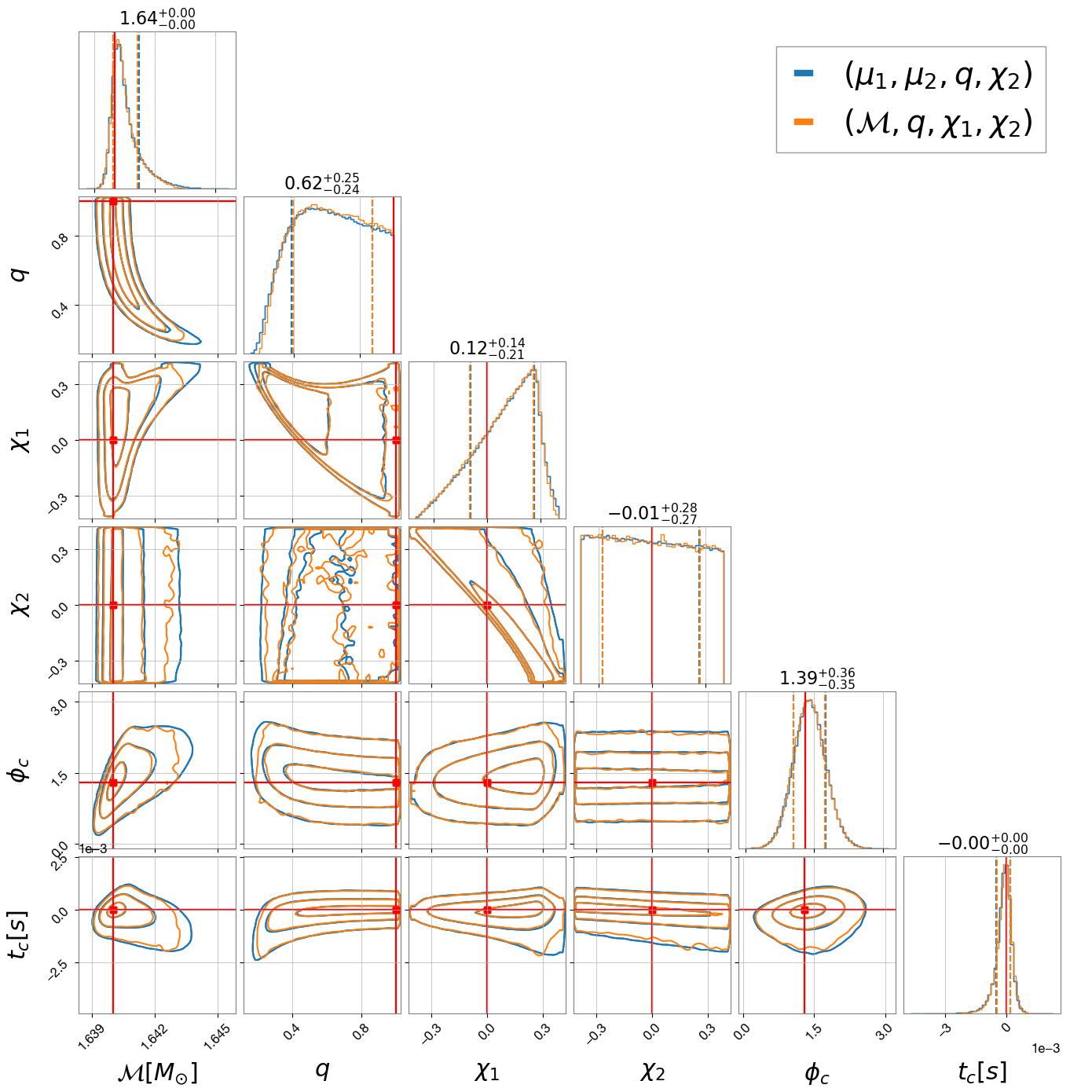}
\caption{Comparison of the estimated posterior distribution with(blue) and without(orange) re-parameterization at $10^8$th iteration in the injection test case \#S2. Dashed lines on each 1-dimensional marginal distribution represent $1\sigma$ region. 3 contours on each 2-dimensional marginal distribution represent $1\sigma$, $2\sigma$ and $3\sigma$ region respectively. The red lines represent the injected value.}
\label{fig:test_compare}
\end{figure*}

\begin{figure*}[t]
\subfloat{%
  \includegraphics[width=.47\linewidth]{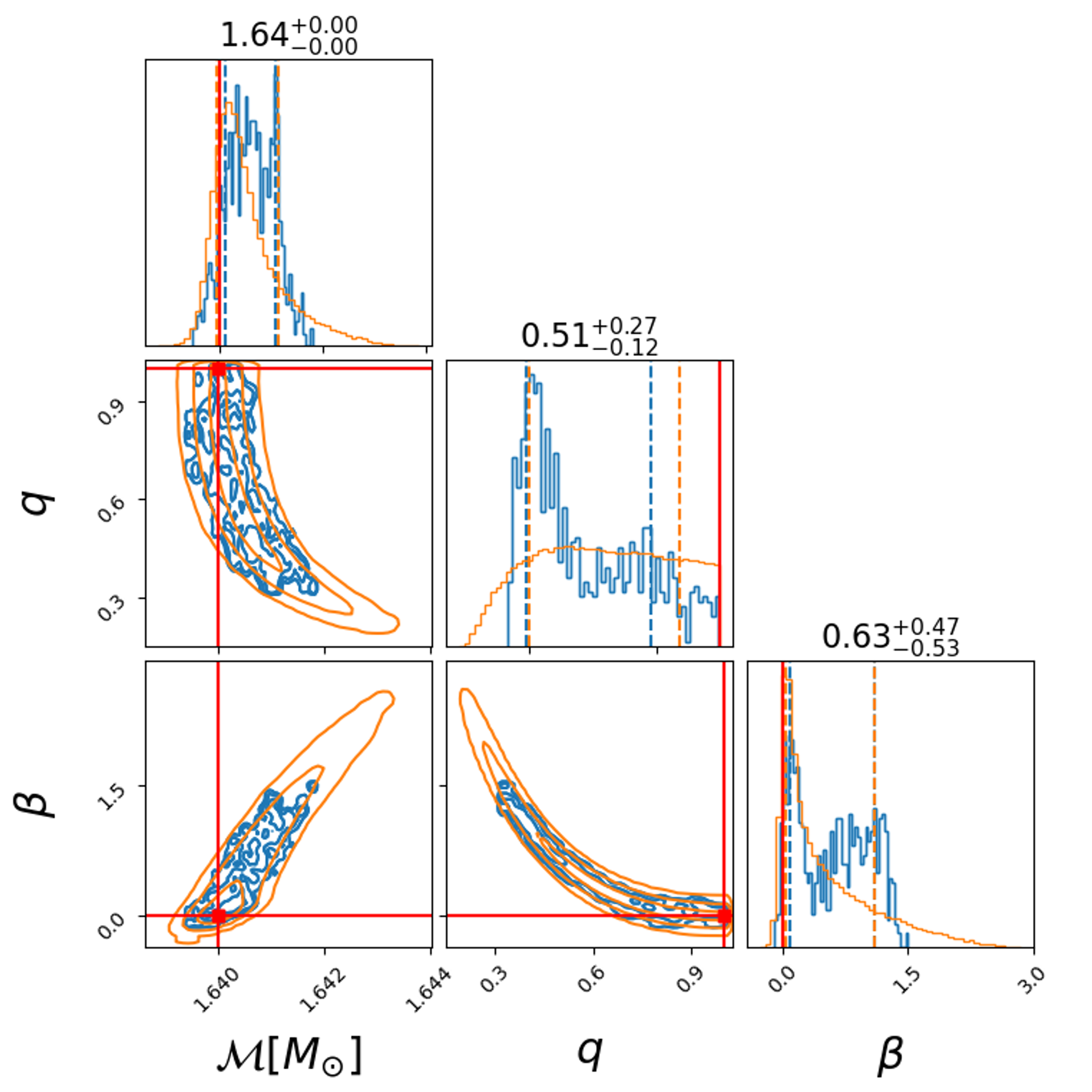}%
}\hfill
\subfloat{%
  \includegraphics[width=.47\linewidth]{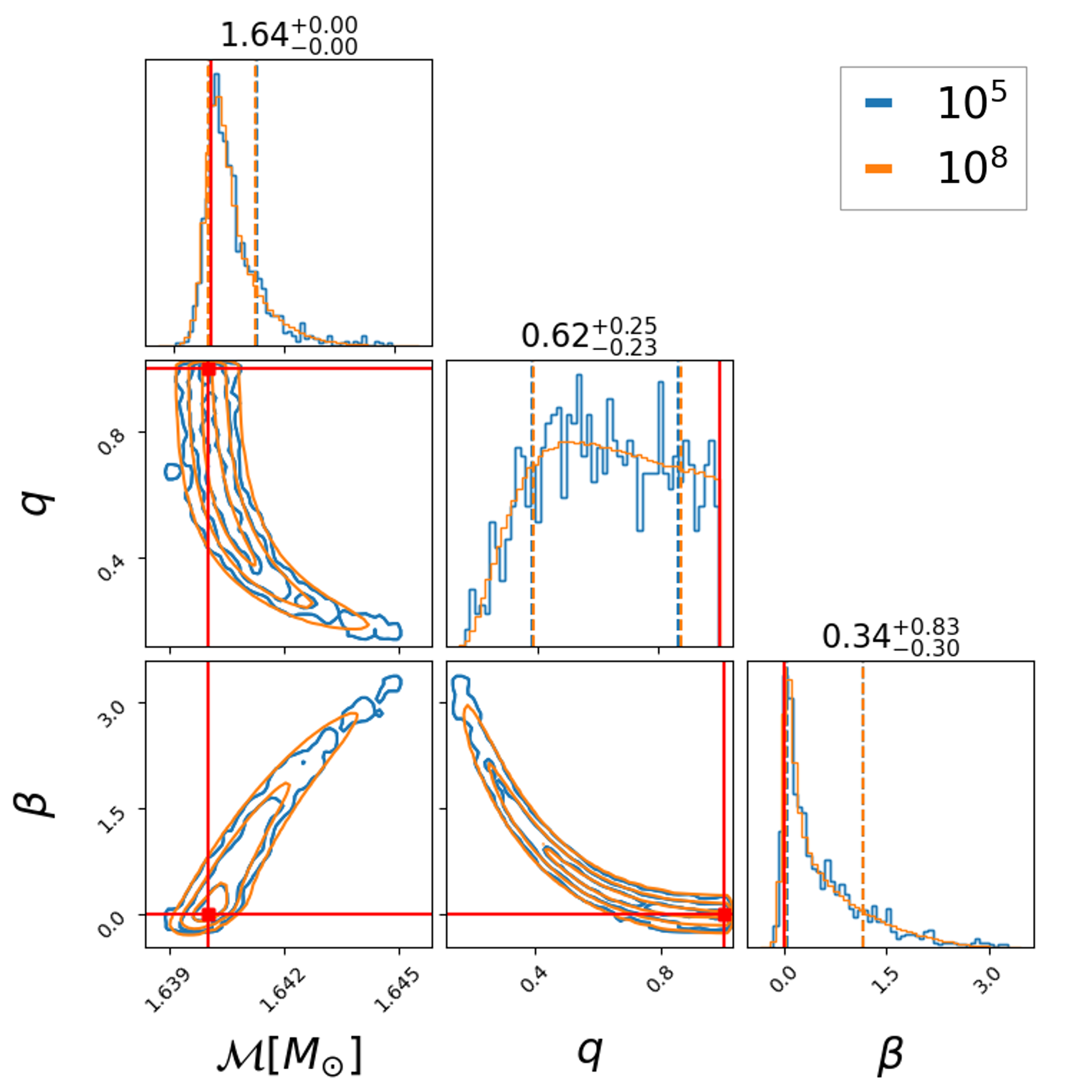}%
}
\caption{Snapshots of the posterior samples at the $10^5$th(blue) and the $10^8$th(orange) iteration in the injection test case \#S2. 
The left and the right panels show samples generated without and with our re-parameterization, respectively. Dashed lines on each 1-dimensional marginal distribution represent $1\sigma$ region. 3 contours on each 2-dimensional marginal distribution represent $1\sigma$, $2\sigma$ and $3\sigma$ region respectively. The red lines represent the injected value.}
\label{fig:conmass_test}
\end{figure*}

Table~\ref{tab:injectionbinary} lists the CBC signals we simulated and the spin prior range for each.
Most of our injected signals are in the BNS mass range, where the parameter estimation analysis is computationally costly and its speedup is necessary.
The massive case \#S7 is to test the effectiveness of our sampling parameters in the BBH mass region, where our choice of $f_\mathrm{max}=2048\,\si{\hertz}$ for calculating $\mu_n$ is not valid.
The lengths of data used for the analyses are $4\,\si{\second}$ for \#S7 and $128\,\si{\second}$ for the other cases. For 128s-data cases, we use Focused, Reduced Order Quadrature(FROQ) technique in the calculation of likelihoods to speed up the estimation \cite{Morisaki_2020}.

Our new set of sampling parameters deals with the correlation of masses and spins, which are measured by the frequency evolution of the signal.
Thus, it is expected to be enough to test them with data from a single detector.
In most of the tests (\#S1--7), the analysis takes into account only the single detector, LIGO-Livingston.
The last case \#M is the exception, where a signal is injected into data of LIGO-Livingston, LIGO-Hanford and Virgo, and all the data are used for parameter estimation.
Their design sensitivities are used for generating Gaussian noise, and also as the PSDs used for calculating likelihood.

Since our method deals with the inefficiency coming from mass--spin degeneracy in the waveform, it is expected to have greater effect when spin parameters have broader prior range.
To confirm this, we test cases with different spin prior range. 
The narrow spin prior $|\chi_{1,2}| < 0.05$ covers expected spins at merger of known BNS\cite{Abbott_2017,Burgay2003}.
The semi-broad spin prior $|\chi_{1,2}| < 0.4$ covers all known neutron stars\cite{Abbott_2017,T.2006}.
For this prior range, we consider 4 injected signals (\#S2--5) to test the effectiveness of our sampling parameters for various mass and spin values.
The broad spin prior $|\chi_{1,2}| < 0.99$ is used as prior agnostic about the astrophysical nature of compact binaries.

In all the cases, the prior is uniform in component masses $m_1$, $m_2$ and spins $\chi_1$, $\chi_2$.
For the sampling to be efficient, the explored range of $\Mc$ is restricted to be $1.63M_\odot \le\mathcal{M}\le 1.65M_\odot$ for the cases except for \#S7, and $25M_\odot \le\mathcal{M}\le 31M_\odot$ for \#S7.
Since the range of $\Mc$ is narrow, when the Jacobian \eqref{eqn:jacobian} is evaluated during the sampling, it is evaluated with the injected value of $\Mc$ rather than its current value.
This approximation makes the prior $p(\mu_1, \mu_2, q, \chi_2)$ dependent only on $q$ and easy to be implemented.
The range of $q$ is restricted to be $q > 1/8$. For the estimation with re-parameterization, we also set constraints on the range of $\mu_1$ and $\mu_2$ directly, to suppress invalid jump proposals. The $\mu_1$ and $\mu_2$ ranges for each case are calculated from the ranges of $(\mathcal{M},q,\chi_1,\chi_2)$ and can be found in the Table~\ref{tab:masspriorrange}.

\begin{table}[b]
    \centering
    \begin{ruledtabular}
    \begin{tabular}{l m{14em}  r l }
        \multicolumn{2}{l}{Parameter} & Injected value & Unit \\
        \hline
        $\phi_c$&Merger phase &1.3 &rad\\
        $t_c$&Merger time &0 & s\\
        $d_L$&Luminosity distance & 200 & Mpc\\
        $\psi$&Polarization angle & 2.659& rad\\
        $\iota$&Orbital inclination & 0.4 & rad\\
        $\alpha$&Right ascension & 1.375& rad\\
        $\delta$&Declination & $-1.2108$& rad\\
    \end{tabular}
    \end{ruledtabular}

    \caption{Injected non-mass--spin parameters. They are shared in all cases.}
    \label{tab:injectiongeometry}
\end{table}

Other than masses and spins, there are 7 parameters characterizing a CBC signal: merger phase, merger time, luminosity distance to the source, polarization angle, the inclination angle between the line of sight and the orbital angular momentum, and right ascension and declination of the source. 
Their injected values are common in all the test cases and listed in the Table~\ref{tab:injectiongeometry}.
Their prior is the standard one used in the analysis by the LIGO-Virgo-KAGRA (See Appendix B of \cite{Abbott_2019}). The range of merger time is $-0.1\,\si{\second}<t_c<0.1\,\si{\second}$.

In the multiple-detector case \#M, we infer all the 11 source parameters.
For the single-detector cases, geometrical parameters such as right ascension and declination are not measurable.
Thus, we infer only 6 of them: masses, spins, merger phase and time, with the other parameter values being fixed to their injected values.

The PSD difference between \#M and the other cases results in the difference in the parameter conversion.
For the single-detector cases, setting $f_\mathrm{min} = 20 \mathrm{Hz}$ and $S_n(f)$ to the design sensitivity of LIGO-Livingston, $\mu_1$ and $\mu_2$ become
\begin{subequations}
\begin{align*}
    \mu_1 &= 0.97320942\psi_1 +0.21269341\psi_2 + 0.08732089\psi_3, \\
    \mu_2 &= -0.22571628\psi_1 +0.81153242\psi_2 + 0.53895018\psi_3.
\end{align*}
\end{subequations}
In the multiple detector case, assuming $\varrho^2_\text{L} \simeq \varrho^2_\text{H} \simeq \varrho^2_\text{V}$ (the optimal SNRs of the injected signal are 10.93, 14.41 and 8.81 for L, H and V respectively), $\mu_1$ and $\mu_2$ become
\begin{subequations}
\begin{align*}
    \mu_1 &= 0.973164\psi_1 +0.21292605\psi_2 + 0.08726009\psi_3, \\
    \mu_2 &= -0.22599821\psi_1 +0.81299212\psi_2 + 0.53662708\psi_3.
\end{align*}
\end{subequations}

\subsection{Results}

In this subsection, we visualize the results, from case \#S2 as an example, and discuss what we can find from them. After that we list the improvement in the estimation efficiency of all cases.

First, we check whether the posterior distribution becomes simple in the new parameter space. In the Figure~\ref{fig:massspin_test}, we visualize the generated samples as 1-dimensional and 2-dimensional marginal distributions using \texttt{corner.py} \cite{corner}. The left group is the distributions in $(\mathcal{M},q,\chi_1,\chi_2)$ subspace, which shows strong correlations of parameters in the 2-dimensional plots. On the other hand, in the right group, $(\mu_1,\mu_2,\chi_1,\chi_2)$ subspace, the posterior distribution shows weak correlation between any two parameters. Especially the $\mu_1$--$\mu_2$ plot shows a hardly correlated distribution, which makes the exploration efficient greatly.

\begin{figure*}
\centering
\includegraphics[width=0.7\linewidth]{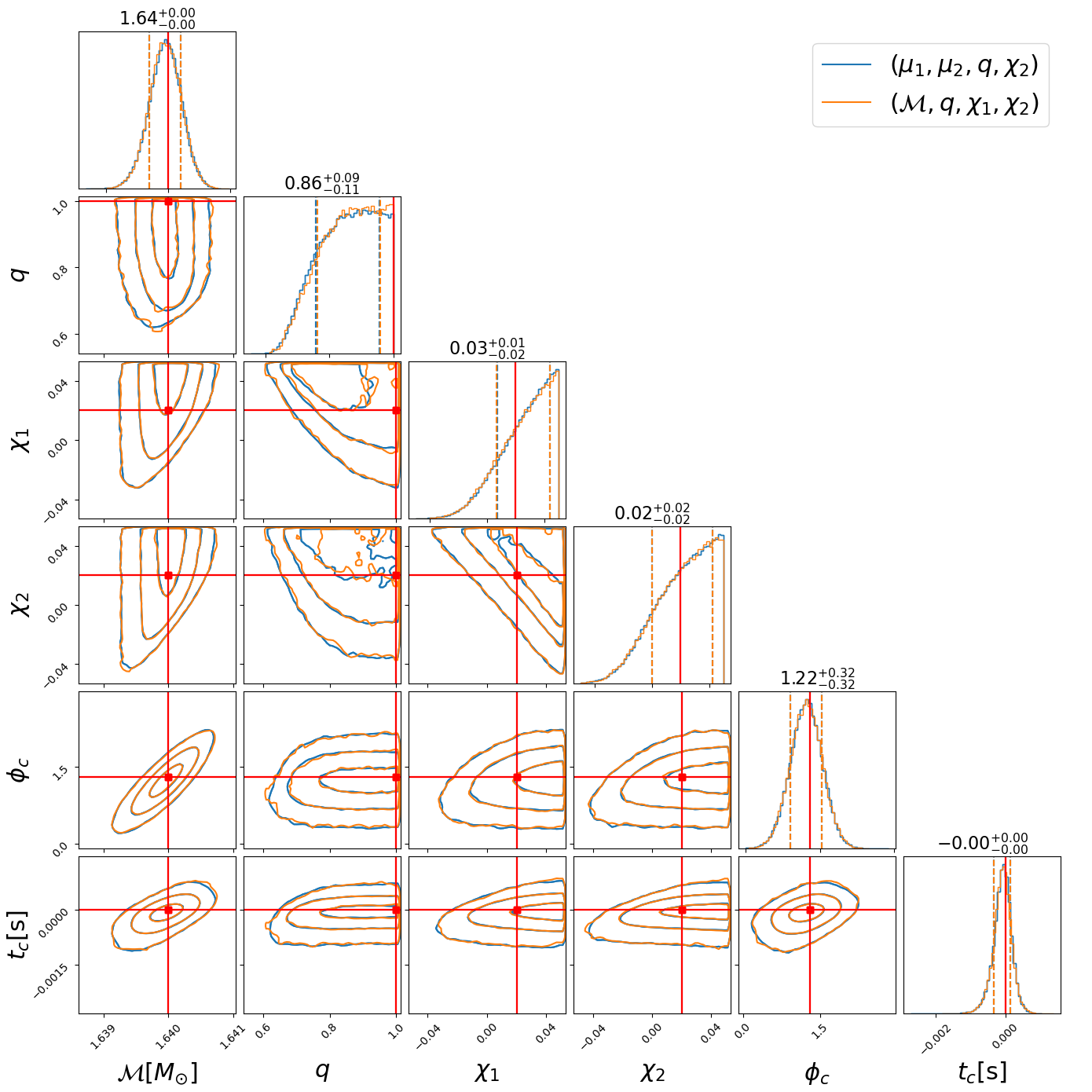}
\caption{Estimated posterior distribution with re-parameterization in the injection test case \#S1. Dashed lines on each 1-dimensional marginal distribution represent $1\sigma$ region. 3 contours on each 2-dimensional marginal distribution represent $1\sigma$, $2\sigma$ and $3\sigma$ region respectively. The red lines represent the injected value.}
\label{fig:na_corner}
\end{figure*}

\begin{figure*}
\centering
\includegraphics[width=0.7\linewidth]{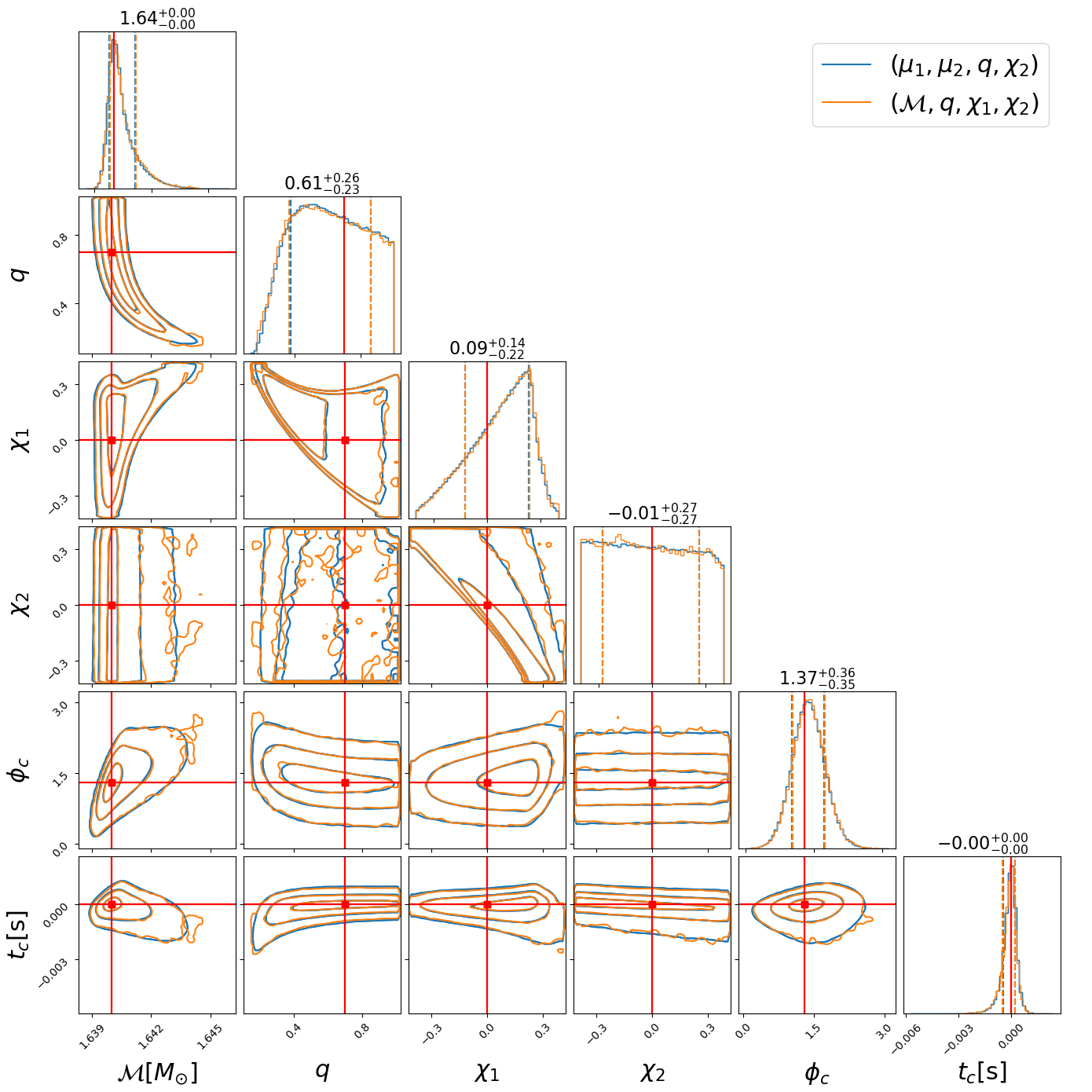}
\caption{Estimated posterior distribution with re-parameterization in the injection test case \#S3. Dashed lines on each 1-dimensional marginal distribution represent $1\sigma$ region. 3 contours on each 2-dimensional marginal distribution represent $1\sigma$, $2\sigma$ and $3\sigma$ region respectively. The red lines represent the injected value.}
\label{fig:zu_corner}
\end{figure*}

\begin{figure*}
\centering
\includegraphics[width=0.7\linewidth]{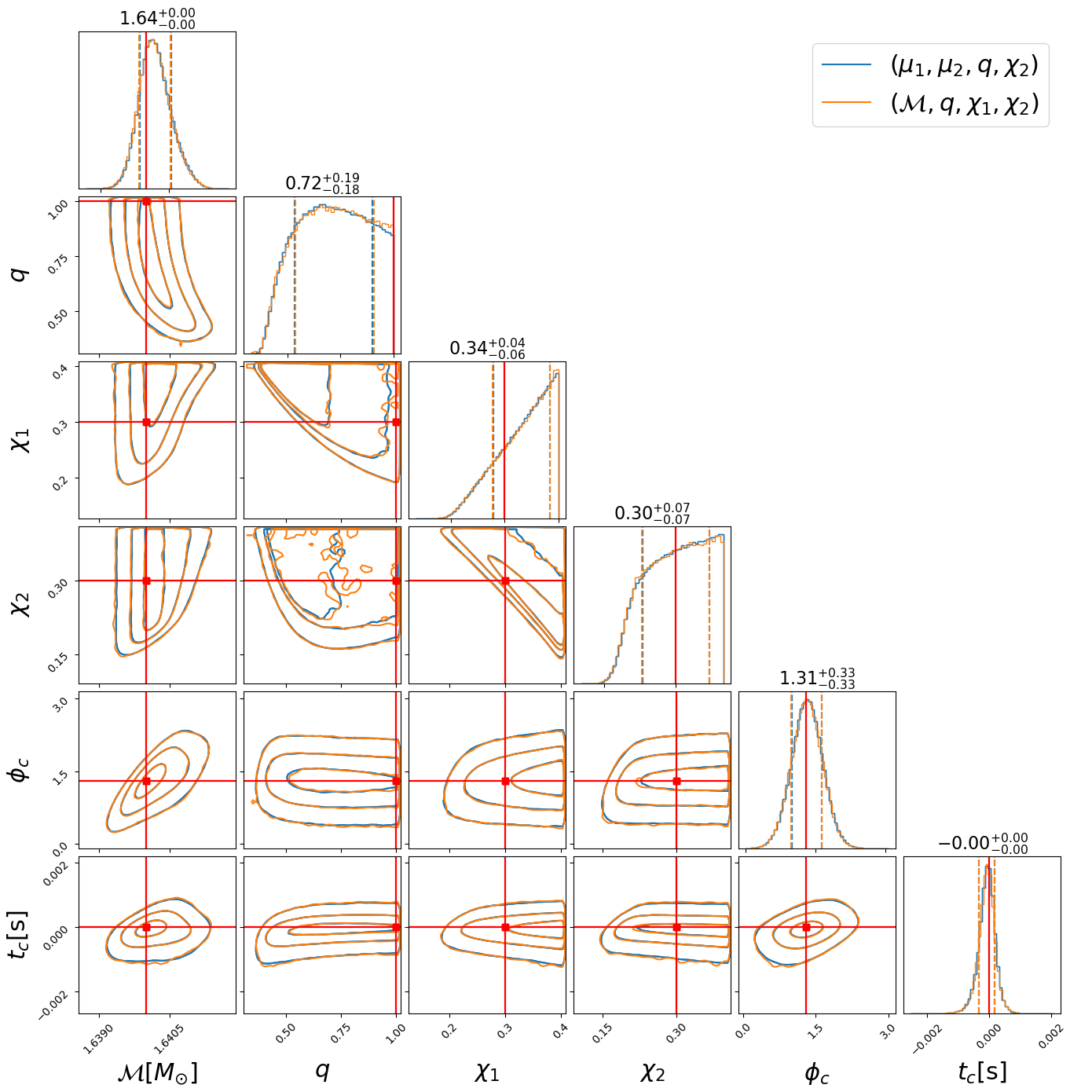}
\caption{Estimated posterior distribution with re-parameterization in the injection test case \#S4. Dashed lines on each 1-dimensional marginal distribution represent $1\sigma$ region. 3 contours on each 2-dimensional marginal distribution represent $1\sigma$, $2\sigma$ and $3\sigma$ region respectively. The red lines represent the injected value.}
\label{fig:he_corner}
\end{figure*}

\begin{figure*}
\centering
\includegraphics[width=0.7\linewidth]{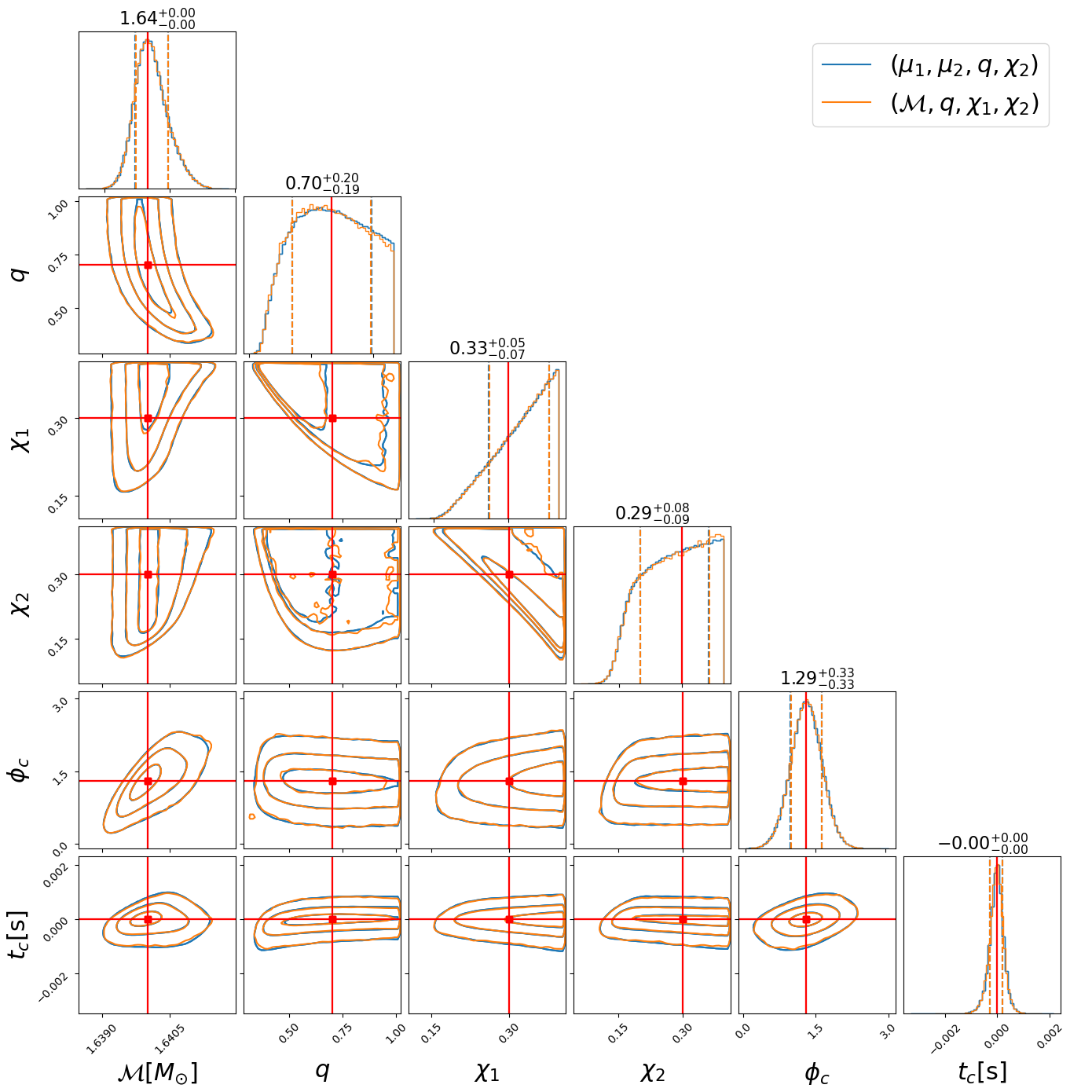}
\caption{Estimated posterior distribution with re-parameterization in the injection test case \#S5. Dashed lines on each 1-dimensional marginal distribution represent $1\sigma$ region. 3 contours on each 2-dimensional marginal distribution represent $1\sigma$, $2\sigma$ and $3\sigma$ region respectively. The red lines represent the injected value.}
\label{fig:hu_corner}
\end{figure*}

\begin{figure*}
\centering
\includegraphics[width=0.7\linewidth]{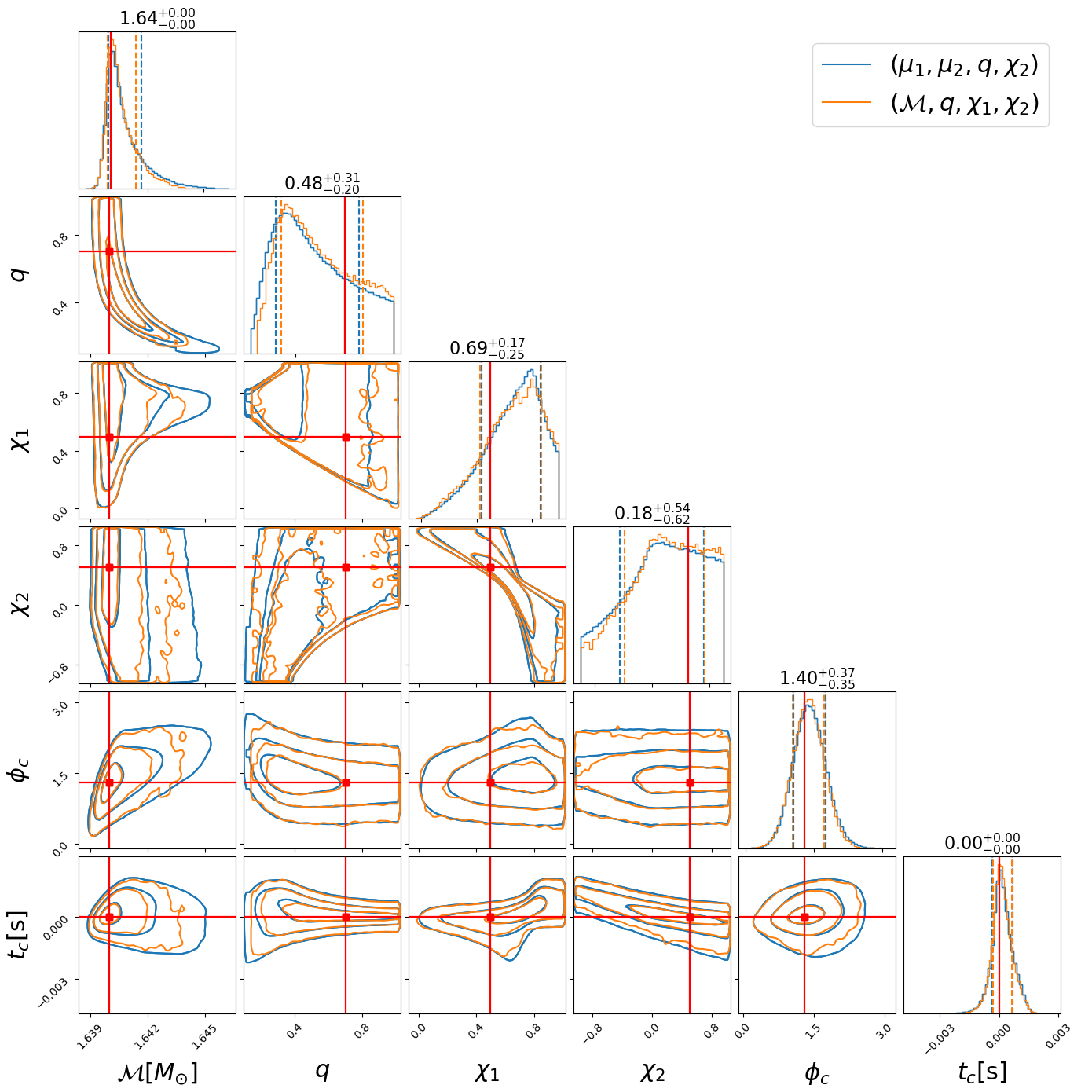}
\caption{Estimated posterior distribution with re-parameterization in the injection test case \#S6. Dashed lines on each 1-dimensional marginal distribution represent $1\sigma$ region. 3 contours on each 2-dimensional marginal distribution represent $1\sigma$, $2\sigma$ and $3\sigma$ region respectively. The red lines represent the injected value.}
\label{fig:br_corner}
\end{figure*}

\begin{figure*}
\centering
\includegraphics[width=0.7\linewidth]{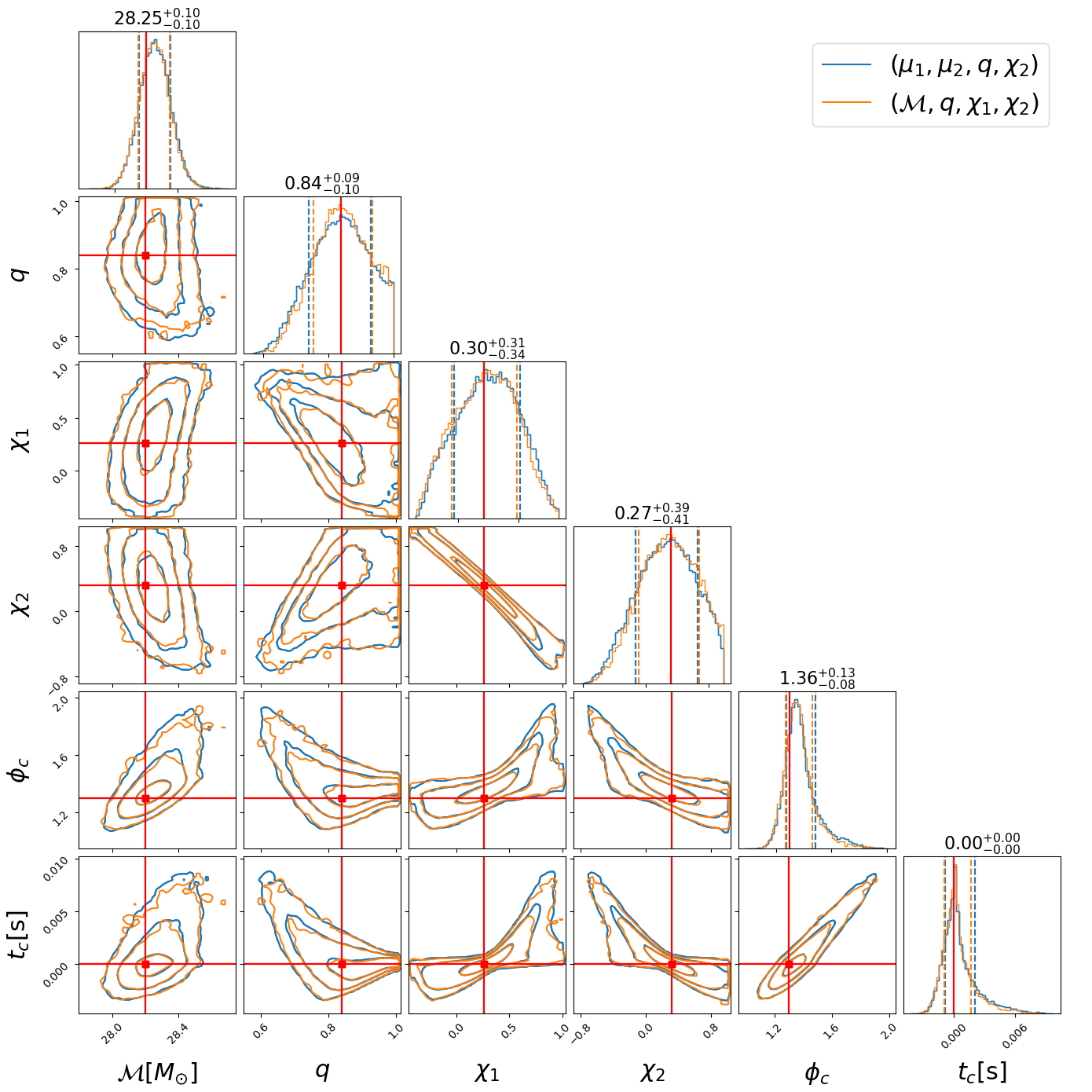}
\caption{Estimated posterior distribution with re-parameterization in the injection test case \#S7. Dashed lines on each 1-dimensional marginal distribution represent $1\sigma$ region. 3 contours on each 2-dimensional marginal distribution represent $1\sigma$, $2\sigma$ and $3\sigma$ region respectively. The red lines represent the injected value.}
\label{fig:bb_corner}
\end{figure*}

\begin{figure*}
\centering
\includegraphics[width=1.0\linewidth]{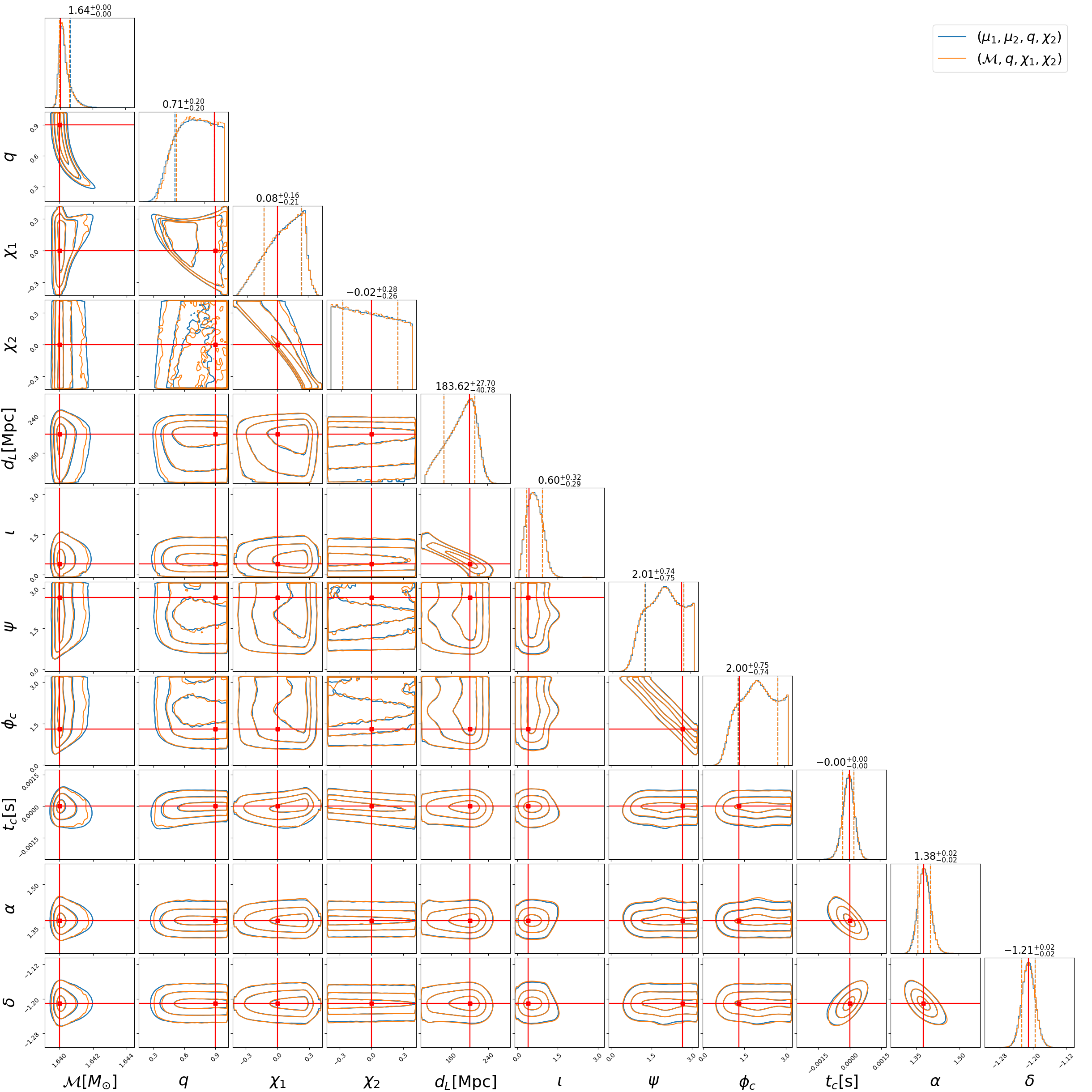}
\caption{Estimated posterior distribution with re-parameterization in the injection test case \#M. Dashed lines on each 1-dimensional marginal distribution represent $1\sigma$ region. 3 contours on each 2-dimensional marginal distribution represent $1\sigma$, $2\sigma$ and $3\sigma$ region respectively. The red lines represent the injected value.}
\label{fig:rl_corner}
\end{figure*}

Next, we check that the estimation results are the same regardless to the sampling parameters. Figure~\ref{fig:test_compare} shows the estimation results with conventional and our new sets of sampling parameters. For better comparison, the density is not plotted in the 2-dimensional plots, and only $1\sigma$, $2\sigma$ and $3\sigma$ contours are plotted. We can see the contours are well consistent between samplings with two different sampling parameters. Also, 1-dimensional distributions clearly agree. For reference, we plot the estimation results comparison from the test cases other than \#S2 too, in Fig~\ref{fig:na_corner}--\ref{fig:rl_corner}.

While the re-parameterization doesn't change the result of estimation, it can reduce the estimation time. Figure~\ref{fig:conmass_test} is a comparison of the distribution of posterior samples generated by the $10^5$th iteration and the $10^8$th iteration. 
Only $\Mc, q, \beta$ of samples are plotted here for simplicity.
In the left plot, which is from the estimation without re-parameterization, the distribution at the $10^5$th iteration is quite different from the converged distribution. In the 2-dimensional plots, we can see an unexplored region. It takes much more iterations to explore the region and converge. 
On the contrary, in the right plot, which is from the estimation with the mass--spin re-parameterization, the samples are already distributed in the entire converged distribution region at the $10^5$th iteration. Especially, 1-dimensional marginal distribution for $\Mc$ and $\beta$ are quite stable at the $10^5$th iteration.

\begin{figure*}[t]
\subfloat{%
  \includegraphics[width=.47\linewidth]{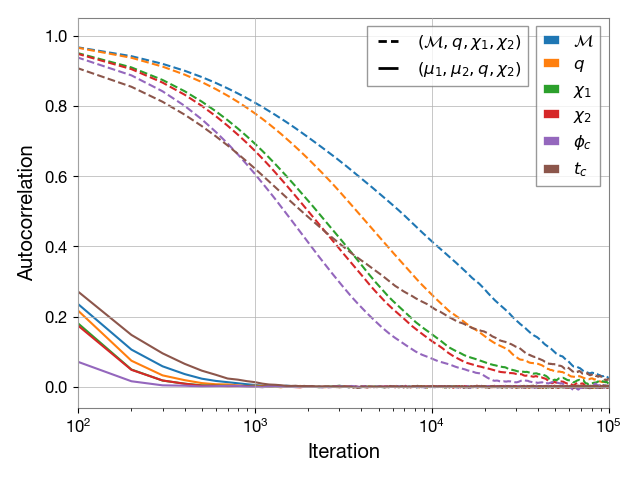}%
}\hfill
\subfloat{%
  \includegraphics[width=.47\linewidth]{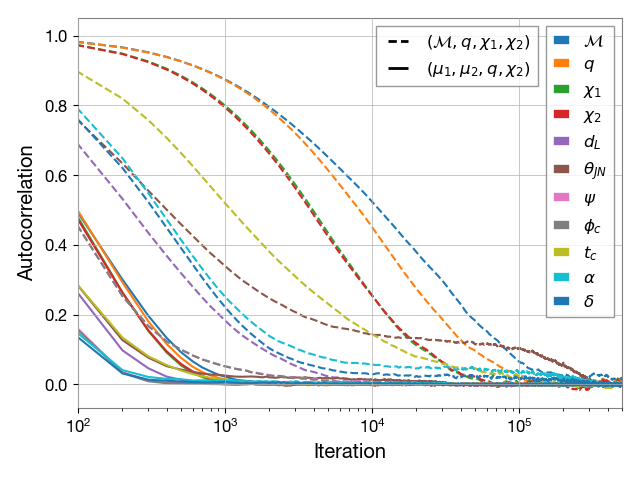}%
}
\caption{Normalized auto-correlations of the generated samples in the two injection test cases, \#S2(left) and \#M(right). The dashed and solid lines represent the sampling using conventional sampling parameters and new parameters, respectively. The symbols of parameters other than masses and spins are introduced in the Table~\ref{tab:injectiongeometry}. }
\label{fig:autocorr_hu}
\end{figure*}

The auto-correlation function of samples can be used to quantify the enhancement of the convergence speed of MCMC algorithms\cite{Sokal1992,Kawashima_1994,aeea2a34a5464c36b24327b48374d96b,Muller-Krumbhaar1973}. In Figure~\ref{fig:autocorr_hu}, each curve represents an normalized auto-correlation function of samples for each parameter. 
The left panel is the result from \#S2, and the right panel is from \#M. The dashed lines are from the estimation without re-parameterization, and the solid lines are from the estimation with re-parameterization. Compared to dashed lines, we can see that the solid lines fall to zero faster, which means that the samples are less correlated with near ones and thus statistically independent samples are generated more frequently. 
Note that, the auto-correlation functions for extrinsic parameters also fall to zero faster with re-parameterization.

The number of iterations to obtain a statistically independent sample is called integrated auto-correlation time(IAT), and can be calculated as
\begin{equation}
    \text{IAT} \approx 1 + 2\sum_{\tau} \hat{c}(\tau)
\end{equation}
where $\tau$ indicates each of iterations and $\hat{c}(\tau)$ is the normalized auto-correlation function.
Therefore we can approximate the speed-up gain by the ratio of IAT, between the estimation with and without the re-parameterization. The $\text{IAT}$ from all cases are listed in Table~\ref{tab:IAT}. A set of samples has different auto-correlation functions and IATs for different parameters. Here, we compare IAT maximized over physical parameters.
We can see that our re-parameterization reduces the IAT by a factor of $\sim10$ for BNS with narrow-spin prior ($|\vec{\chi}|<0.05$) and $\sim100$ for broad-spin prior ($|\vec{\chi}|<0.99$). Even for the massive BBH case, the IAT becomes about 1/3, by using our re-parameterization. 

Although additional time is taken for parameter conversion in each iteration with our method, its effect to the total estimation time is small. To check this, we compare the estimation time directly. In the case \#M, it takes 17 minutes to get 1000 statistically independent samples with re-parameterization on ICRR common computer system \cite{icrr}(CPU: Intel Xeon Gold 6230(2.1GHz)), while 41 hours are needed without re-parameterization. In all cases the total estimation time is reduced.

\begin{table}[h!]
\centering
\begin{ruledtabular}
\begin{tabular}{l r r r}
Case & $(\mathcal{M},q,\chi_1,\chi_2)$ & $(\mu_1,\mu_2,q,\chi_2)$ & Ratio \\
\hline
\#S1 & 427 &47.8&8.93\\
\#S2 & $3.44\times 10^4$ & 250& 138\\
\#S3 & $7.68\times 10^4$ & 276& 278\\
\#S4 & $1.46\times 10^3$ & 50.4 & 29.0\\
\#S5 & $2.00\times 10^3$ & 53.9 & 37.1\\
\#S6 & $1.10\times 10^5$ & 717 & 153 \\
\#S7 & $1.45\times 10^4$ & $5.04\times 10^3$ & 2.88\\
\#M & $6.17 \times 10^4$ & 401 & 154 \\
\end{tabular}
\end{ruledtabular}
\caption{The comparison of maximal(over physical parameters) IAT values estimated from the samples.}
\label{tab:IAT}
\end{table}

\section{Conclusions} \label{sec:conclusions}

In this paper we have introduced a new set of mass--spin parameters for aligned-spin compact binary inspiral waveform, which makes the posterior distribution simple and therefore the estimation efficient. To inspect its effect on the sampling efficiency, we performed parameter estimation runs on simulated signals using the new set of mass--spin sampling parameters. In all test cases, the new set of parameters improves the efficiency of the sampling process. Especially the improvement is remarkable for the analysis of binary neutron star signals with a broad prior range of spins, where the effects of mass--spin correlations are significant. Quantitatively, the speed-up gain in the analysis of binary neutron star signals is $\sim10$ for narrow-spin prior ($|\vec{\chi}|<0.05$), $\sim10$--$100$ for semi-broad-spin prior ($|\vec{\chi}|<0.4$), and $\sim100$ for broad-spin prior ($|\vec{\chi}|<0.99$). 

The results are case-dependent, thus they have to be understood carefully. In the tests, we adopted single component adaptive Metropolis and adaptive Metropolis jump proposals. The choice is natural, but there could be other jump proposals that make the sampling with complicated posterior more efficient. Using those proposals may reduce the improvement of our method since it settles the inefficiency in a different way. On the other hand, it can also reduce the integrated auto-correlation time with our method, so using both should be a good choice to optimize the estimation.

We can combine other parameter estimation techniques with our re-parameterization. Especially, since our method reduces the number of likelihood evaluations, it could be a nice duo with methods that cut down the single likelihood evaluation time, such as the focused reduced order quadrature technique. Our method also can be used with parallel tempering, which can reduce the estimation time additionally.

In the narrow spin prior case(\#S1) and the massive case(\#S7), the posterior distribution already has a relatively simple form in the usual mass--spin parameter space. Even in these cases, our re-parameterization improves the sampling process, and at least does not worsen the sampling. This fact, with the improvement in the multiple detectors case, shows our method could be applied to the actual observation  comprehensively.

In this paper, we only consider a binary system whose spins are aligned with its orbital angular momentum. If the spins are misaligned with the orbital angular momentum, the precession of the orbital plane is induced, and the amplitude and phase of signal are modulated \cite{Apostolatos:1994mx}. Since it can break the degeneracy between distance and orbital inclination angle, and improve the accuracy of source localization \cite{Vitale:2018wlg, Tsutsui:2020bem}, rapid parameter estimation taking into account the precession effects can be helpful for multi-messenger observations. Even in this case, the frequency evolution of signal is predominantly determined by masses and spin components along the orbital angular momentum, and our re-parameterization may still make the posterior distribution simple and the analysis more efficient. We leave the extension of our method to precessing binary systems for a future work.
\\

\begin{acknowledgments}

We thank Kyohei Kawaguchi, Tatsuya Narikawa, Nami Uchikata, Bin-Hua Hsieh and Takashi Kato for useful discussions and comments. This work was supported by NSF PHY-1912649 (S.M.), MEXT, JSPS Leading-edge Research Infrastructure Program, JSPS Grant-in-Aid for Specially Promoted Research 26000005, JSPS Grant-in-Aid for Scientific Research on Innovative Areas 2905: JP17H06358, JP17H06361, JP16H02183 and JP17H06364, JSPS Core-to-Core Program A. Advanced Research Networks, JSPS Grant-in-Aid for Scientific Research (S) 17H06133 and 15H00787, the joint research program of the Institute for Cosmic Ray Research, 
the cooperative research program of the Institute of Statistical Mathematics, National Research Foundation (NRF) and Computing Infrastructure Project of KISTI-GSDC in Korea, Academia Sinica (AS), AS Grid Center (ASGC) and the Ministry of Science and Technology (MoST) in Taiwan under grants including AS-CDA-105-M06, Advanced Technology Center (ATC) of NAOJ, Mechanical Engineering Center of KEK, the LIGO project, and the Virgo project.

\end{acknowledgments}

\bibliographystyle{apsrev4-1}
\bibliography{mass_spin_reparam}

\end{document}